\documentclass[aps,prl,twocolumn,superscriptaddress,showpacs]{revtex4-2}
\usepackage{amsmath}
\usepackage[dvips,xdvi]{graphicx}
\usepackage{amssymb}
\usepackage{epsfig}
\usepackage{xcolor}
\usepackage[colorlinks=true,citecolor=black,urlcolor=black,linkcolor=black]{hyperref}

\begin{document}
	\title{Engineering Dynamical Phase Diagrams with Driven Lattices in Spinor Gases}
	\author{J. O. Austin-Harris}
	\author{Z. N. Hardesty-Shaw}
	\affiliation{Department of Physics, Oklahoma State University, Stillwater, Oklahoma 74078, USA}
	\author{Q. Guan}
	\affiliation{Homer L. Dodge Department of Physics and Astronomy, The University of Oklahoma, Norman,
		Oklahoma 73019, USA}
	\affiliation{Center for Quantum Research and Technology, The University of Oklahoma, Norman, Oklahoma 73019, USA}
	\affiliation{Department of Physics and Astronomy, Washington State University, Pullman, WA 99164, USA}
	
	\author{C. Binegar}
	\affiliation{Department of Physics, Oklahoma State University, Stillwater, Oklahoma 74078, USA}
	
	\author{D. Blume}
	\affiliation{Homer L. Dodge Department of Physics and Astronomy, The University of Oklahoma, Norman,
		Oklahoma 73019, USA}
	\affiliation{Center for Quantum Research and Technology, The University of Oklahoma, Norman, Oklahoma 73019, USA}
	
	\author{R.~J. Lewis-Swan}
	\email{lewisswan@ou.edu} \affiliation{Homer L. Dodge Department of Physics and Astronomy, The University of Oklahoma,
		Norman, Oklahoma 73019, USA} \affiliation{Center for Quantum Research and Technology, The University of Oklahoma, Norman,
		Oklahoma 73019, USA}
	\author{Y. Liu}
	\email{yingmei.liu@okstate.edu}
	\affiliation{Department of Physics, Oklahoma State University, Stillwater, Oklahoma 74078, USA}
	\date{\today}
	\begin{abstract}
		We experimentally demonstrate that well-designed driven lattices are versatile tools to
		simultaneously tune multiple key parameters (namely spin-dependent interactions, spinor phase, and Zeeman energy) for manipulating phase diagrams of spinor gases with negligible heating and atom losses. This opens a new avenue for studying dynamical phase transitions in engineered Hamiltonians. The driven lattice creates additional separatrices in phase space at driving-frequency-determined locations, with progressively narrower separatrices at higher Zeeman energies due to modulation-induced higher harmonics. The vastly expanded range of magnetic fields at which significant spin dynamics occur and improved sensitivities at higher harmonics represent a step towards quantum sensing with ultracold gases.	
	\end{abstract}
	\maketitle
	
	The high degree of controllability available in lattice-confined spinor Bose-Einstein condensates (BECs) provides an ideal platform for quantum simulation of a wide range of interesting phenomena, spanning from nonequilibrium dynamics and dynamical phase transitions to the production of massively entangled spin singlets with immediate applications in quantum enhanced sensing~\cite{Stamper2013,Ueda2012,Zach1,Zach2,Chen2019,Austin1,Austin2,Lichao2018,Romano2004,Zhang2005,Chang2005,Kronjager2006,
		Black2007,Yingmei2009,Pechkis2013,He2015,Lichao2014,Lichao2015,Zhang2005,Chang2005,Jiang2016,Jiang2014,Guan2021,Zhou2023,Gerbier2021}. These and many other spinor phenomena are induced by interconversions 
	of multiple spin components and therefore determined by the competition of the quadratic Zeeman energy $q$ and the spin-dependent
	interaction $c_2$~\cite{Stamper2013,Ueda2012,Zach1,Zach2,Chen2019,Austin1,Austin2,Lichao2018,Romano2004,Zhang2005,Chang2005,
		Kronjager2006,Black2007,Yingmei2009,Pechkis2013,He2015,Lichao2014,Lichao2015,Zhang2005,Chang2005,Jiang2016,Jiang2014}. Due to 
	intrinsically small values of $c_2$ in many atomic species, studies of spinor physics and its applications have mainly been 
	restricted to a regime with small values of $q$ comparable to $c_2$ (e.g., weak magnetic fields), motivating an extensive search 
	for methods of manipulating these quantities~\cite{Zhang2005,Chang2005,Kronjager2006,Black2007,Yingmei2009,Pechkis2013,Lichao2015,He2015,Zach1,Chen2019,Austin1,Austin2,Jiang2016,Lichao2014}. Common methods of tuning $c_2$ include changing the atomic density, confining atoms into static optical lattices, or altering scattering lengths near Feshbach
	resonances, however, these techniques are limited by heating and atom losses for substantial changes~\cite{
		Stamper2013,Ueda2012,Lichao2015,Zach1,Chin2010,Knoop2011PRA}. Manipulation of $q$ can be accomplished via dressing fields induced by, e.g., off-resonant microwave pulses or linearly polarized off-resonant laser beams~\cite{Stamper2013,Ueda2012,Lichao2014,Jiang2014,
		Santos2007,Leslie2009,Gerbier2006, Bookjans2011,Lichao2014,Jiang2014}. Similar to the issues with $c_2$ manipulation, 
	substantial changes remain limited due to excess heating and atom losses, while the spinor physics remains restricted to relatively small net $q$~\cite{Lichao2014,Jiang2014}. 
	
	\begin{figure*}[tb]
		\includegraphics[width=176mm]{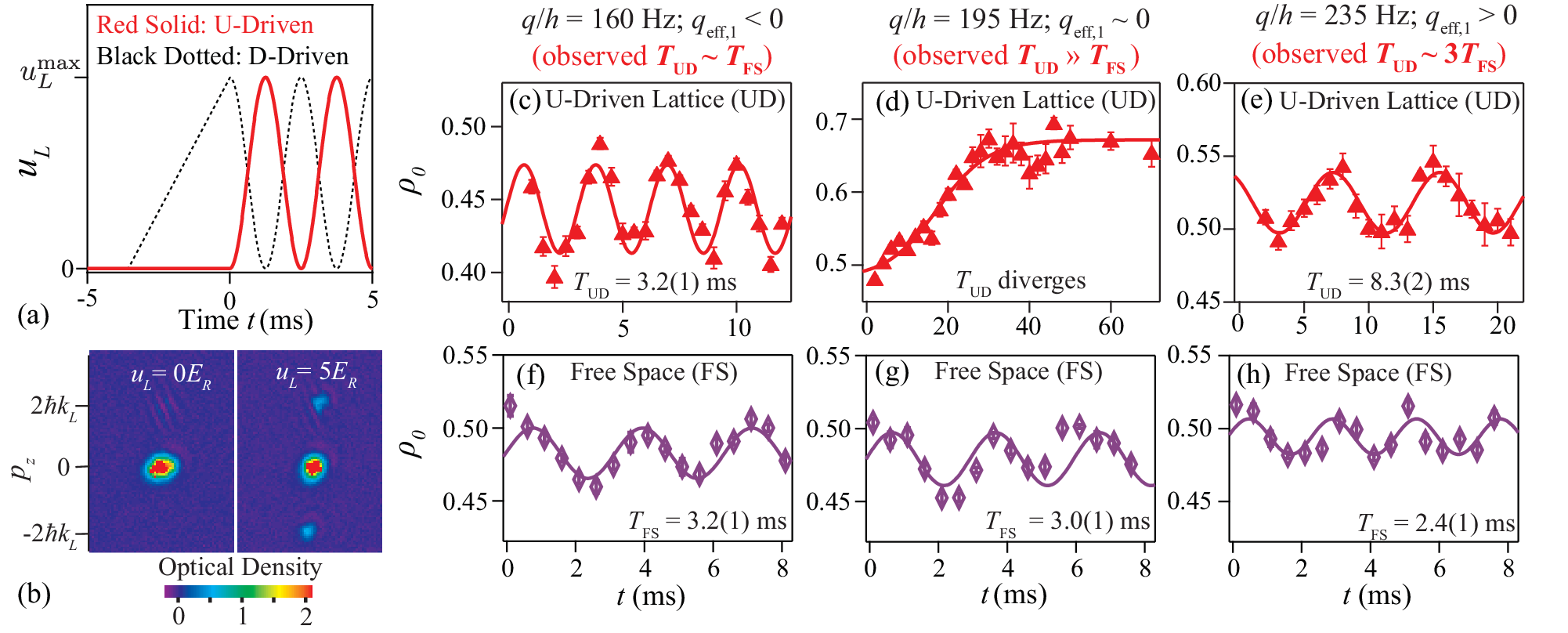}
		\caption{ (a) Red solid (black dotted) line displays the timing of U-Driven (D-Driven) lattice sequences. (b) Sample TOF images taken via U-Driven sequences at the minimum (left) and maximum (right) lattice depth, showing atoms occupying different momentum $p$ states. Here $\hbar$ is the reduced Planck constant and $k_L$ is the lattice vector. (c-e) Triangles display experimentally observed spin oscillations near $q^*/h\approx f/2=200~\mathrm{Hz}$ via U-Driven sequences when  $q/h$ equals (c) $160~\mathrm{Hz}$, (d) $195~\mathrm{Hz}$, and (e)	$235~\mathrm{Hz}$. (f-h) Similar to panels (c) to (e) but taken in free space. Solid lines in panels (c)-(h) are sinusoidal or sigmoidal fits~\cite{SM}.} \label{Fig1}
	\end{figure*}
	
	In this paper, we present a versatile method for manipulating dynamical phase diagrams of spinor gases by sinusoidally driving the depth of a one-dimensional (1D) optical lattice at a frequency $f$ much larger than the typical scale of spin-dependent interactions $c_2$. Our data demonstrate that well-designed driven lattices can simultaneously tune multiple key parameters that determine spinor physics, including spin-dependent interactions, spinor phase, and $q$, with negligible heating and atom losses. Prior studies of spin dynamics in free space (no lattices) have identified a diverging oscillation period at a critical quadratic Zeeman energy $q^*$, which is associated with the crossing of a separatrix in the underlying classical phase space and marks a dynamical phase transition~\cite{Marino2022,Robert2021}. We observe that the driven lattices create additional dynamical critical points at $q^*$ approximately symmetric about $q/h = f/2$ (where $h$ is the Planck constant), with progressively narrower critical regions occurring whenever $q/h$ is an integer multiple of $f/2$ due to higher harmonics generated by the lattice modulation. These observations are understood by theoretical calculations based on a dynamical single spatial model approximation (dSMA)~\cite{Zach1,SM}.  The vastly expanded range of magnetic fields at which significant spin dynamics occur and the improved sensitivities at higher harmonics represent a step towards applications of ultracold atoms for precise quantum sensing. 
	
	In each experimental cycle, we prepare an initial state of $\rho_0\approx 0.5$, $\theta=0$, and magnetization $M=\rho_{1}-\rho_{-1}=0$ with an $F$=1 BEC of up to $10^5$ sodium atoms at $t=0$ and a desired $q$. Here $\rho_{m_F}$ is the fractional population in the $|F=1,m_F\rangle$ state and $\theta$ is the relative phase among the three spin states \cite{SM}. The depth of a 1D lattice is then sinusoidally driven at $f=400~\mathrm{Hz}$ between $0$ and the maximum depth $u_L^{\mathrm{max}}$, i.e., $u_L(t)=(u_L^{\mathrm{max}}/2)(1+\cos(2\pi f t-\phi))$ with $\phi=0$ ($\phi=\pi$) for D-Driven (U-Driven) sequences (see Fig.~\ref{Fig1}(a) and Supplemental Materials~\cite{SM}). We monitor spin dynamics via spin-resolved imaging after time of flight (TOF) ballistic expansion~\cite{Lichao2014,Chen2019,SM}.
	
	The driven lattice induces oscillations of the momentum distributions (Fig.~\ref{Fig1}(b)) and a periodic spin-dependent interaction given by $c_2(t) = \mathcal{G}_0 + \sum_{k=1}^{\infty} \mathcal{G}_{k} \cos(k 2\pi f t - \phi_k)$. 
	The index $k$ arises because $c_2(t)$ is only approximately sinusoidal, the higher harmonics ($k>1$) therefore have nonzero contributions to the dynamics~\cite{SM}. Here, $\mathcal{G}_{0}$ ($\mathcal{G}_{k}$) is a positive interaction strength controlling the elastic (inelastic) spin-preserving (spin-mixing) collisions~\cite{SM}. The phase $\phi_k$ is determined by the specific driven lattice sequence, for example, $\phi_1 = \phi$ and $\phi_2 = 2\phi + \pi$ for our experimental sequences~\cite{SM}. In this work $c_2/h\sim 25~\mathrm{Hz}$ in free space and for all driven lattice sequences we set $u_L^{\rm max}=5E_R$, where $E_R$ is the recoil energy, yielding $\mathcal{G}_1/\mathcal{G}_0\sim0.2$ with $\mathcal{G}_{0}/h\sim 30~\mathrm{Hz}$ and $\mathcal{G}_1/h\sim 5~\mathrm{Hz}$~\cite{SM}. For our system, where $M=0$ and $hf\gg\mathcal{G}_0$ and $\mathcal{G}_1$, we apply a dSMA to describe the spin dynamics in a frame rotating with the modulated interaction using a mean-field Hamiltonian (see Supplemental Materials~\cite{SM}), 
	\begin{equation}\label{eqn:HamMF}
		H_{\mathrm{mf},j}=\rho_0(1-\rho_0) \left(\mathcal{G}_0+ \frac{\mathcal{G}_{j}}{2}\cos\theta_{\mathrm{eff},j}\right) + q_{\mathrm{eff},j}(1-\rho_0),
	\end{equation}
	where the $j$-th order effective quadratic Zeeman energy is $q_{\mathrm{eff},j}= q - j h f/2$ and the effective spinor phase is $\theta_{\mathrm{eff},j}=\theta+\phi_j$. Equation~\ref{eqn:HamMF} bears similarities to the established static SMA (sSMA) model of spinor BECs in free space, but enables the independent tuning of spin-dependent interactions via $\mathcal{G}_0$ and $\mathcal{G}_1$ while the interaction can only be tuned by $c_2$ in free space~\cite{SM}.  Crucially, the other key quantities, $\theta$ and $q$, are also replaced by effective quantities $\theta_{\mathrm{eff},j}$ and $q_{\mathrm{eff},j}$ defined by the driven lattice. 
	
	\begin{figure*}[t]
		\includegraphics[width=176mm]{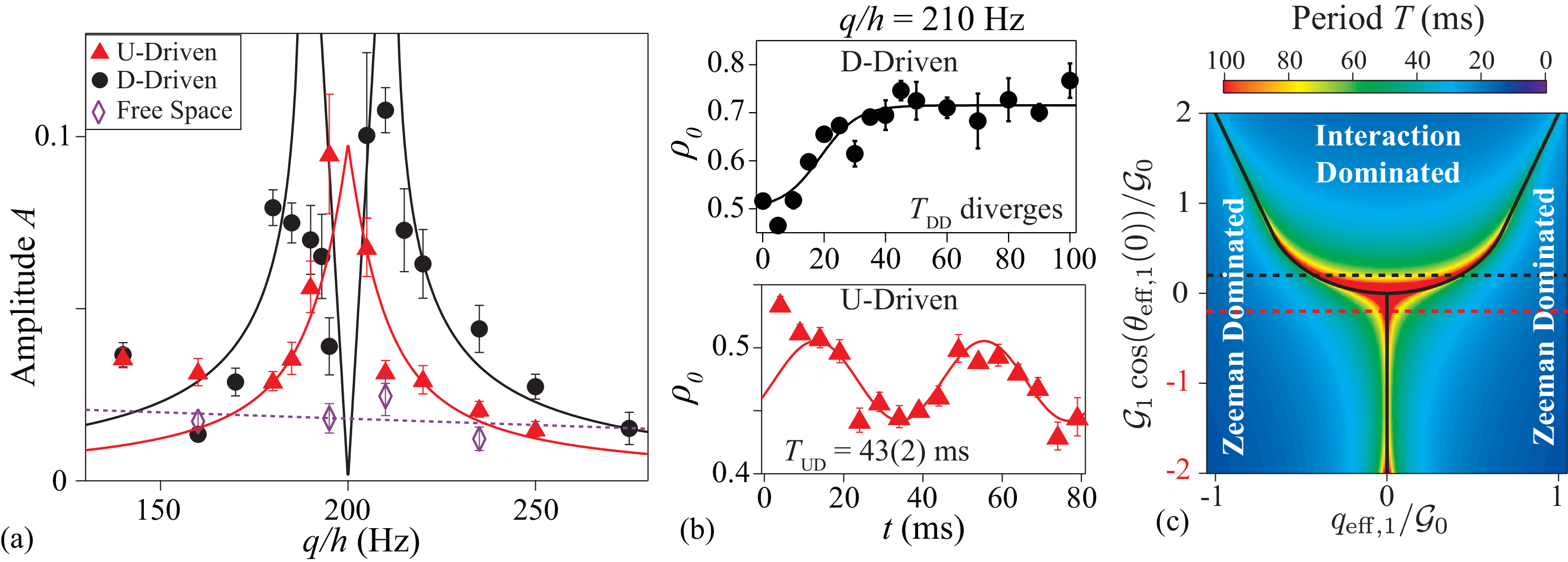}
		\caption{(a) Triangles (circles) [diamonds] display spin oscillation amplitudes observed via U-Driven (D-Driven) [free space] sequences near $q/h=f/2=200~\mathrm{Hz}$. Solid lines are dSMA predictions, while the dashed line is a linear fit to guide the eye. (b) Drastically different spin dynamics observed via D-Driven (upper graph) and U-Driven (lower graph) sequences at $q/h=210~\mathrm{Hz}$. Solid lines are sinusoidal or sigmoidal fits. (c) A phase diagram of $H_{\mathrm{mf},1}$ (see Eq.~1) for our system demonstrates the dynamical critical points (black solid lines) for driven lattice systems are determined by $\mathcal{G}_1\cos(\theta_{\mathrm{eff},1}(0))/\mathcal{G}_0$. The black (red) dashed line marks $\mathcal{G}_1\cos(\theta_{\mathrm{eff},1}(0))/\mathcal{G}_0=0.2$ $(=-0.2)$ for the D-Driven (U-Driven) data studied in this paper.} \label{Fig2}
	\end{figure*}
	
	Typical spin dynamics observed via U-Driven sequences near $q/h=f/2=200~\mathrm{Hz}$ are displayed in Figs.~\ref{Fig1}(c)-\ref{Fig1}(e). In the displayed time traces, the period $T$ and amplitude $A$ of spin oscillations vary from a relatively small value of $T=3.2(1)~\mathrm{ms}$ and $A=0.032(4)$ at $q/h=160~\mathrm{Hz}$ (Fig.~\ref{Fig1}(c)) to a diverging period at $q/h=195~\mathrm{Hz}$ (Fig.~\ref{Fig1}(d)) before returning to a small value of $T=8.3(2)~\mathrm{ms}$ and $A=0.021(2)$ as $q$ is further increased to $q/h=235~\mathrm{Hz}$ (Fig.~\ref{Fig1}(e)). For the same $q$ region in free space, spin oscillations are drastically different with roughly constant $T\sim 3~\mathrm{ms}$ and $A\sim 0.015$ (Figs.~\ref{Fig1}(f-h)). 
	
\begin{figure*}[tbh]
		\includegraphics[width=176mm]{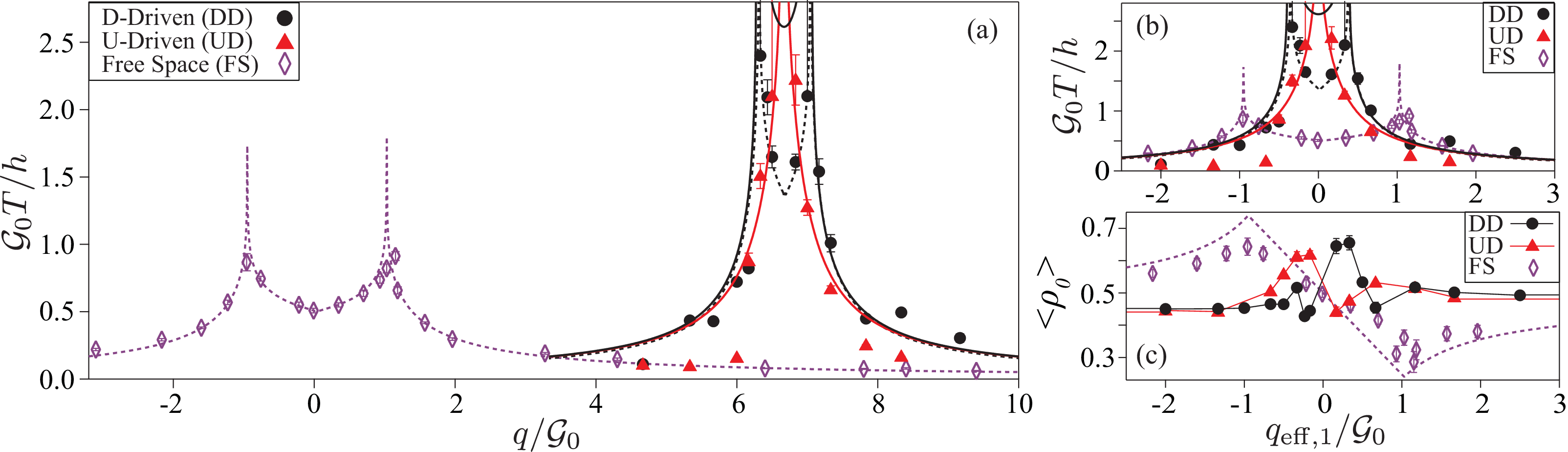}
		\caption{(a) Spin oscillation period observed via U-Driven (triangles) and D-Driven (circles) sequences compared to free space data taken under a combination of magnetic and microwave dressing fields (diamonds). Some of the free-space data are adapted from our prior work~\cite{Lichao2014}. Dashed lines are established free space sSMA predictions for
			our initial state (with the black dashed line shifted by $q/h=200~\mathrm{Hz}$), while solid lines are dSMA predictions. 
			(b) Similar to (a) but plotted against $q_{\mathrm{eff},1}/h=q/h-f/2$. (c) Similar to (b) but plots center of oscillations $\langle\rho_0\rangle$. For free space data, due to the absence of driven lattices, $f=0$ and $\mathcal{G}_{\mathrm{0}}$ equals the free space $c_2$.}
		\label{Fig3}
	\end{figure*}

	Figure~\ref{Fig2}(a) compares spin oscillations observed in free space and via the different driven lattice sequences as a function of $q$. With U-Driven sequences, we observe a single peak centered at a dynamical critical point $q^*/h\approx f/2=200~\mathrm{Hz}$ (see triangles in Fig.~\ref{Fig2}(a)), which is characterized by anharmonic oscillations of divergent period separating regions of harmonic spin oscillations for $q<q^*$ and $q>q^*$. In contrast, the observed oscillation period and amplitude in free space remain small and slowly decrease with $q$ (see diamonds in Fig.~\ref{Fig2}(a)) consistent with the sSMA predictions (as further elaborated in Fig.~\ref{Fig3}(a))~\cite{Lichao2014,Stamper2013}. Figure~\ref{Fig1} and Fig.~\ref{Fig2}(a) therefore demonstrate that U-Driven sequences can induce an additional dynamical critical point at a driving-frequency-determined location of $q^*/h\approx f/2$, corresponding to the predicted driven lattice induced resonance at $q_{\mathrm{eff},1}=0$.

	Our driven lattice system can also precisely tune the effective initial phase $\theta_{\mathrm{eff},j}(0)$ between 0 and $2\pi$ with a step of $4\times10^{-4}$ radians.  This resolution could be further improved with more precise hardware.  Modulation of the initial phase, although it has been much less explored in the literature due to technical challenges, can drastically alter the system behavior as exemplified by different time traces taken at a same $q$ via U-Driven and D-Driven sequences (see Fig.~\ref{Fig2}(b)). This is further studied in Fig.~\ref{Fig2}(a): in contrast to U-Driven sequences with only one dynamical critical point, two distinct peaks are observed via D-Driven sequences (see black circles in Fig.~\ref{Fig2}(a)). These observations agree with the theoretical driven lattice phase diagram derived from Eq.~1 (see Fig.~\ref{Fig2}(c) and solid lines in Fig.~\ref{Fig2}(a)), in which the Zeeman-dominated regime with a characteristic period $T \approx h/2q_{\mathrm{eff},1}$ and the interaction-dominated regime with a $T$ set by the interactions $\mathcal{G}_0$ and $\mathcal{G}_1$ are split by a critical $q^*$ where $T$ diverges~\cite{SM}. When $\theta_{\mathrm{eff},1}(0)=\pi$ and thus $\mathcal{G}_1\cos(\theta_{\mathrm{eff},1}(0))/\mathcal{G}_0=-\mathcal{G}_1/\mathcal{G}_0<0$, as realized by U-Driven sequences (red dashed line in Fig.~\ref{Fig2}(c)), the system displays a single $q^*$ at $q_{\mathrm{eff},1}=q-hf/2=0$. In contrast, when $\theta_{\mathrm{eff},1}(0)=0$ and thus $\mathcal{G}_1\cos(\theta_{\mathrm{eff},1}(0))/\mathcal{G}_0=\mathcal{G}_1/\mathcal{G}_0>0$, as realized by D-Driven sequences (black dashed line in Fig.~\ref{Fig2}(c)), the system displays two $q^*$. 	
	
	Normalizing $q$ and $T$ by $\mathcal{G}_0$ enables direct comparisons of the spin dynamics observed in driven and undriven systems, as shown in Fig.~\ref{Fig3} with triangles (circles) [diamonds] displaying two other characteristics ($T$ and center $\langle\rho_0\rangle$) of spin oscillations taken with U-Driven (D-Driven) [free space] sequences. Here $\langle\rho_0\rangle$ is extracted from averaging $\rho_0(t)$ over a given time trace. Figure~\ref{Fig3}(a) clearly demonstrates that both driven lattice sequences display significant spin dynamics centered at $q/h\approx f/2=200~\mathrm{Hz}$ ($q/\mathcal{G}_0\sim 6.7$) which do not appear in free space. To examine this effect in more detail, the same data are instead plotted against $q_{\mathrm{eff},1}$ in Fig.~\ref{Fig3}(b) revealing stark similarities between the dynamics observed in free space and via D-Driven sequences, with the critical regions created by D-Driven lattices (in free space) centered at $q_{\mathrm{eff},1}^*\sim\pm 0.37 \mathcal{G}_0=\pm h\times 11~\mathrm{Hz}$ ($q_{\rm eff,1}^*\sim \pm \mathcal{G}_0$) for the same initial state with $\theta_{\mathrm{eff},1}(0)=0$. This agrees with the dSMA predicted critical regions that are symmetrically located at $q_{\mathrm{eff},1}^*\sim \pm 0.42 \mathcal{G}_0$ for our D-Driven system where $\mathcal{G}_1\cos(\theta_{\mathrm{eff},1}(0))/\mathcal{G}_0\sim 0.2$ (black dashed line in Fig.~\ref{Fig2}(c)). Therefore, upon initiating or stopping the D-Driven protocol for $q$ between the two symmetric $q^*$, e.g., $189~\mathrm{Hz}<q/h<211~\mathrm{Hz}~(-0.37<q_{\mathrm{eff},1}/\mathcal{G}_0<0.37)$, our system realizes dynamical phase transitions between the Zeeman-dominated regime and the interaction-dominated regime (see Fig.~\ref{Fig2}(a) and Fig.~\ref{Fig3}(b)).

	The observed period for U-Driven sequences confirms a single dynamical critical point at $q_{\mathrm{eff},1}\approx 0$ (see Fig.~\ref{Fig3}(b)), which is similar to the predicted behavior of spinor BECs in free space with the same phase $\theta_{\mathrm{eff},1}(0)=\pi$. The separatrix in U-driven sequences is additionally marked by a sudden jump in $\langle \rho_0 \rangle$, while for free space the two $q^*$ are characterized by extrema of $\langle \rho_0 \rangle$ with a gradual transition between the maximum and minimum values (see Fig.~\ref{Fig3}(c)). Figure~\ref{Fig2}(a) and Fig.~\ref{Fig3} therefore demonstrate that the spinor phase can be conveniently tuned by altering the phase $\phi$ of the lattice modulation, consistent with the prediction of  Eq.~\ref{eqn:HamMF}.
	
	While the dSMA provides qualitative insight into the emergence of new dynamical critical points, it does not quantitatively capture several aspects of our experimental observations. For data taken via D-Driven sequences, the observed periods in the interaction-dominated regime are only qualitatively captured by the dSMA. Oddly these data quantitatively agree with a shifted sSMA prediction (black dashed lines in Figs.~\ref{Fig3}(a-b)) with $c_2/h=11~\mathrm{Hz}$, i.e., much smaller than the estimated free space $c_2/h\sim 25~\mathrm{Hz}$, despite not having a physical motivation. Additionally, data taken via U-Driven sequences display much smaller periods than predicted when $q$ is relatively far from $q^*$. Finally, the observed $\langle \rho_0 \rangle$ of both driven lattice sequences displays behavior reflected about $q_{\mathrm{eff},1}=0$ from the predictions. A complete understanding of the observed rich physics requires further study with more sophisticated models beyond the dSMA (see Supplemental Materials~\cite{SM}). 
	
\begin{figure}[b]
		\includegraphics[width=86mm]{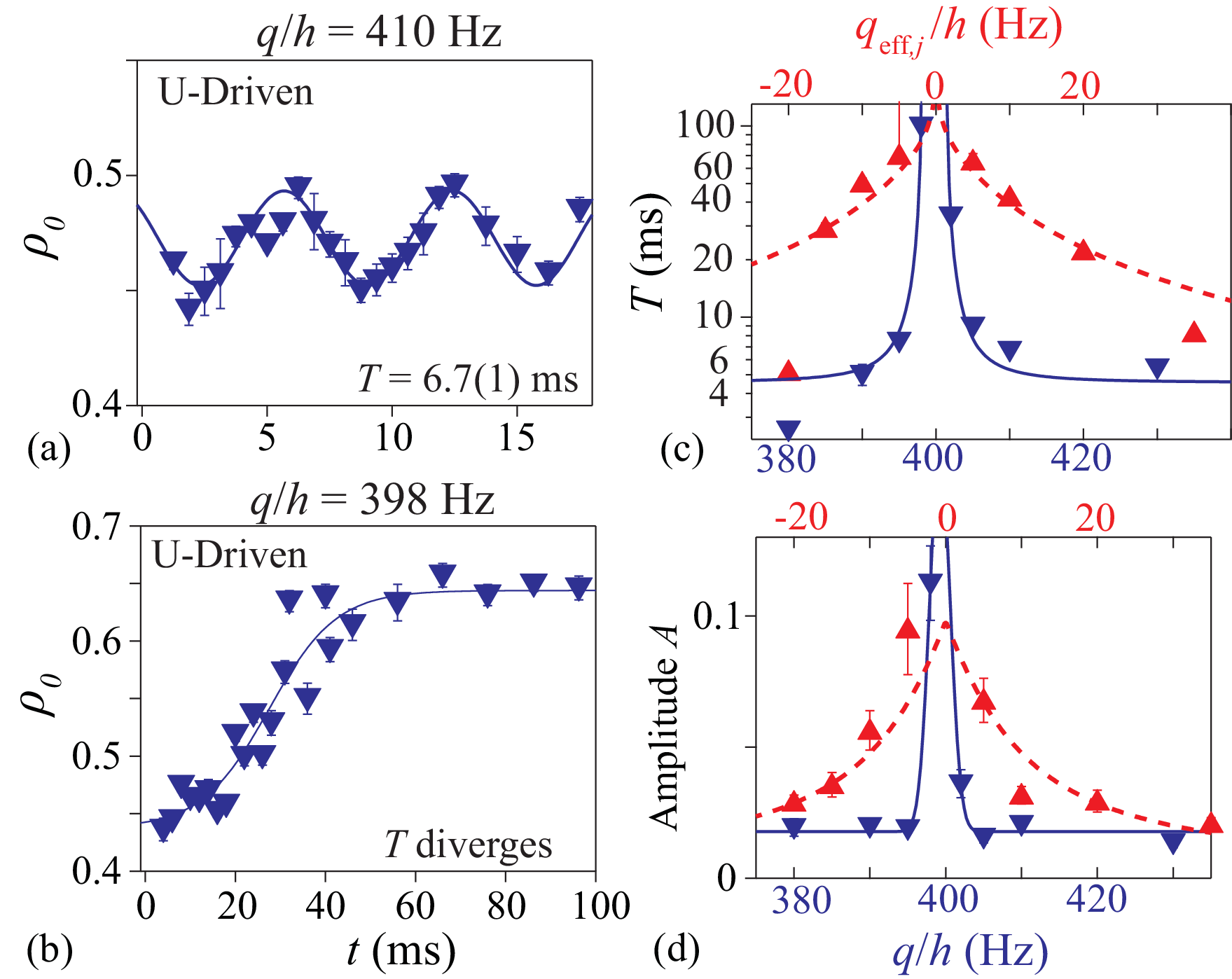}
		\caption{ Time traces taken when $q/h$ equals (a) $410~\mathrm{Hz}$ and (b) $398~\mathrm{Hz}$
			demonstrating a critical region at a higher harmonic with U-Driven sequences. Solid lines are sinusoidal or sigmoidal fits. (c-d) Blue (red) triangles display the oscillation period (c) and amplitude (d) observed via U-Driven sequences near $q^*/h\approx 400~\mathrm{Hz}$ ($200~\mathrm{Hz}$) which corresponds to $q_{\mathrm{eff},2}=0$ ($q_{\mathrm{eff},1}=0$). The top (bottom) axis is applied to both datasets (blue triangles).  Red dashed (blue solid) lines are dSMA predictions (Lorentzian fits to guide the eye)~\cite{SM}.} \label{Fig4}
	\end{figure}

	As we only approximately drive $c_2(t)$ sinusoidally~\cite{SM}, the higher harmonics have nonzero contributions to the Fourier decomposition of $c_2(t)$ and therefore contribute to the spin dynamics. As these higher harmonics have much weaker contributions to $c_2(t)$~\cite{SM}, progressively narrower critical regions occur in the driven lattice system whenever $q/h$ is an integer multiple of $f/2$, as demonstrated by two time traces taken near $q^*/h\approx f=400~\mathrm{Hz}$ with U-Driven sequences in Fig.~\ref{Fig4}. The time trace taken at $q/h=410~\mathrm{Hz}$ (Fig.~\ref{Fig4}(a)) can be compared to the time trace taken at $q/h=210~\mathrm{Hz}$ (Fig.~\ref{Fig2}(b)) to reveal that for the same $q_{\mathrm{eff},j}$, the data taken near $q^*/h\approx f$ ($q_{\mathrm{eff},2}= 0$) has a smaller period and amplitude than data taken near $q^*/h\approx f/2$ ($q_{\mathrm{eff},1}= 0$). However, as shown by the time trace taken at $q/h=398~\mathrm{Hz}$ in Fig. 4(b), the response near the dynamical critical point is unmistakable. We use the oscillation period (amplitude) to map out the critical region near $q^*/h\approx f=400~\mathrm{Hz}$ with U-Driven sequences in Fig.~\ref{Fig4}(c) (\ref{Fig4}(d)) observing similar, although narrower, signatures to the U-Driven data taken near $q^*/h\approx f/2=200~\mathrm{Hz}$. For example, the width at half maximum of the observed separatrix is around $5~\mathrm{Hz}$ ($20~\mathrm{Hz}$) near $q^*/h\approx 400~\mathrm{Hz}$ ($200~\mathrm{Hz}$). Critical regions at the higher harmonics, for example at $q^*/h\approx f$, can therefore be more sensitive to magnetic field strengths and may have immediate applications to precise quantum sensing. For example, when using U-Driven sequences, due to the sharp peak in oscillation amplitude at $q^*/h\approx f$, a handful of points at relatively long hold times in the lattice (around 50~ms in our system) should be sufficient to locate $q^*$ by properly scanning $f$. As $q^*$ in a U-Driven system is determined solely by $f$, it can be used as a precise system-independent probe. 	
	
	We have introduced a versatile method to simultaneously tune multiple key parameters (namely spin-dependent interactions, $\theta_{\mathrm{eff},j}$, and $q_{\mathrm{eff}, j}$) by sinusoidally driving the lattice depth, enabling experimental realizations of dynamical phase transitions with individually tunable elastic and inelastic collisions. This opens a new avenue to the study of engineered Hamiltonians~\cite{Fujimoto2019,Zhang2023,Evrard2019,Li2019,Jose2023}. This driven lattice method enables the observation of spin dynamics at large magnetic fields while overcoming some challenges, for example the detrimental atom losses, associated with other widely used techniques. The contributions of higher harmonics are observed as increasingly more sensitive critical regions occurring when $q/h$ is an integer multiple of $f/2$.  This can be important for the exploitation of critical spin dynamics for sensing magnetic fields over a wide working range. 
	
	\begin{acknowledgments}
		\noindent{\it Acknowledgments --} D.~B. acknowledges support by the National Science Foundation (NSF) through grant No.
		PHY-2110158. R.~J. L-S. acknowledges support by NSF through Grant No. PHY-2110052 and the Dodge Family College of Arts and
		Sciences at the University of Oklahoma. J.~O.~A-H., Z.~N.~H-S., C.~B. and Y.~L. acknowledge support by the Noble
		Foundation and NSF through Grant No. PHY-2207777. 
	\end{acknowledgments}


\begin{thebibliography}{100}%
		\section{References}
		\makeatletter
		\providecommand \@ifxundefined [1]{%
			\@ifx{#1\undefined}
		}%
		\providecommand \@ifnum [1]{%
			\ifnum #1\expandafter \@firstoftwo
			\else \expandafter \@secondoftwo
			\fi
		}%
		\providecommand \@ifx [1]{%
			\ifx #1\expandafter \@firstoftwo
			\else \expandafter \@secondoftwo
			\fi
		}%
		\providecommand \natexlab [1]{#1}%
		\providecommand \enquote  [1]{``#1''}%
		\providecommand \bibnamefont  [1]{#1}%
		\providecommand \bibfnamefont [1]{#1}%
		\providecommand \citenamefont [1]{#1}%
		\providecommand \href@noop [0]{\@secondoftwo}%
		\providecommand \href [0]{\begingroup \@sanitize@url \@href}%
		\providecommand \@href[1]{\@@startlink{#1}\@@href}%
		\providecommand \@@href[1]{\endgroup#1\@@endlink}%
		\providecommand \@sanitize@url [0]{\catcode `\\12\catcode `\$12\catcode
			`\&12\catcode `\#12\catcode `\^12\catcode `\_12\catcode `\%12\relax}%
		\providecommand \@@startlink[1]{}%
		\providecommand \@@endlink[0]{}%
		\providecommand \url  [0]{\begingroup\@sanitize@url \@url }%
		\providecommand \@url [1]{\endgroup\@href {#1}{\urlprefix }}%
		\providecommand \urlprefix  [0]{URL }%
		\providecommand \Eprint [0]{\href }%
		\providecommand \doibase [0]{http://dx.doi.org/}%
		\providecommand \selectlanguage [0]{\@gobble}%
		\providecommand \bibinfo  [0]{\@secondoftwo}%
		\providecommand \bibfield  [0]{\@secondoftwo}%
		\providecommand \translation [1]{[#1]}%
		\providecommand \BibitemOpen [0]{}%
		\providecommand \bibitemStop [0]{}%
		\providecommand \bibitemNoStop [0]{.\EOS\space}%
		\providecommand \EOS [0]{\spacefactor3000\relax}%
		\providecommand \BibitemShut  [1]{\csname bibitem#1\endcsname}%
		\let\auto@bib@innerbib\@empty
		%</preamble>
		{
			
			\bibitem{Ueda2012}%
			\bibinfo{author}{Kawaguchi, Y.} \& \bibinfo {author}{Ueda, M.},
			\newblock \bibinfo {title}{Spinor {B}ose--{E}instein
				condensates}.
			\newblock \emph{\bibinfo{journal}{Phys. Rep.}}
			\textbf {\bibinfo {volume}{520}},\ \bibinfo {pages} {253} (\bibinfo {year} {2012}).
			
			
			\bibitem{Stamper2013}%
			\bibinfo{author}{Stamper-Kurn, D.~M.} \& \bibinfo{author}{Ueda, M.},
			\newblock \bibinfo{title}{Spinor {B}ose gases: symmetries, magnetism, and
				quantum dynamics}.
			\newblock \emph{\bibinfo{journal}{Rev. Mod. Phys.}}
			\textbf{\bibinfo{volume}{85}}, \bibinfo{pages}{1191} (\bibinfo{year}{2013}).
			
			\bibitem{Romano2004}%
			\bibinfo {author} {Romano, D.~ R.} \& \bibinfo {author} {de Passos, E.~J.~V.}, 
			\newblock \bibinfo{title} {Population and phase dynamics of $F= 1$ spinor condensates in an external magnetic field}.
			\newblock \emph{\bibinfo{journal}{Phys. Rev. A}} \textbf{\bibinfo{volume}{70}}, \bibinfo{pages}{043614} (\bibinfo{year}{2004}).
			
			\bibitem{Zach2}%
			\bibinfo {author} {Hardesty-Shaw, Z.~N.}, \bibinfo {author} {Guan, Q.},
			\bibinfo {author} {Austin-Harris, J.~O.}, \bibinfo
			{author} {Blume, D.}, \bibinfo {author}
			{Lewis-Swan, R.~J.}, \& \bibinfo {author}
			{Liu, Y.}, \bibinfo{title} {Nonlinear multi-state tunneling dynamics in a spinor Bose-Einstein condensate}. \emph{\bibinfo{journal} {arXiv}}\textbf{: 2306.05877} (\bibinfo{year}{2023}).
			
			\bibitem{Chen2019}%
			\bibinfo{author}{Chen, Z.}, \bibinfo{author}{Tang, T.}, \bibinfo{author}{Austin, J.}, \bibinfo{author}{Shaw, Z.}, \bibinfo{author}{Zhao, L.}, \& \bibinfo{author}{Liu, Y.},
			\newblock \bibinfo{title}{Quantum quench and nonequilibrium dynamics in lattice-confined spinor condensates}.
			\newblock \emph{\bibinfo{journal}{Phys. Rev. Lett.}}
			\textbf{\bibinfo{volume}{123}}, \bibinfo{pages}{113002}
			(\bibinfo{year}{2019}).
			
			
			\bibitem{Zach1}%
			\bibinfo {author} {Hardesty-Shaw, Z.~N.}, \bibinfo {author} {Guan, Q.},
			\bibinfo {author} {Austin, J.~O.}, \bibinfo
			{author} {Blume, D.}, \bibinfo {author}
			{Lewis-Swan, R.~J.}, \& \bibinfo {author}
			{Liu, Y.}, \bibinfo{title} {Quench-induced nonequilibrium dynamics of spinor gases in a moving
				lattice}. \emph{\bibinfo{journal} {Phys. Rev. A}} \textbf{\bibinfo {volume}
				{107}}, \bibinfo{pages}{053311} (\bibinfo{year}{2023}).
			
			\bibitem{Austin2}%
			\bibinfo{author}{Austin, J.~O.}, \bibinfo{author}{Shaw, Z.~N.},
			\bibinfo{author}{Chen, Z.}, \bibinfo{author}{Mahmud, K.~W.}, \&
			\bibinfo{author}{Liu, Y.},
			\newblock \bibinfo{title}{Manipulating atom-number distributions and detecting
				spatial distributions in lattice-confined spinor gases}.
			\newblock \emph{\bibinfo{journal}{Phys. Rev. A}}
			\textbf{\bibinfo{volume}{104}}, \bibinfo{pages}{L041304}
			(\bibinfo{year}{2021}).
			
			\bibitem{Austin1}%
			\bibinfo{author}{Austin, J.~O.}, \bibinfo{author}{Chen, Z.},
			\bibinfo{author}{Shaw, Z.~N.}, \bibinfo{author}{Mahmud, K.~W.}, \&
			\bibinfo{author}{Liu, Y.},
			\newblock \bibinfo{title}{Quantum critical dynamics in a spinor {Hubbard} model
				quantum simulator}.
			\newblock \emph{\bibinfo{journal}{Comm. Phys.}} \textbf{\bibinfo{volume}{4}},
			\bibinfo{pages}{61} (\bibinfo{year}{2021}).
			
			
			\bibitem{Zhang2005}%
			\bibinfo{author}{Zhang, W.}, \bibinfo{author}{Zhou, D.~L.}, \bibinfo{author}{Chang, M.-S.}, \bibinfo{author}{Chapman, M.~S.}, \& \bibinfo{author}{You, L.},
			\newblock \bibinfo{title}{Coherent spin mixing dynamics in a spin-1 atomic condensate}.
			\newblock \emph{\bibinfo{journal}{Phys. Rev. A}}
			\textbf{\bibinfo{volume}{72}}, \bibinfo{pages}{013602} (\bibinfo{year}{2005}).
			
			\bibitem{Chang2005}%
			\bibinfo{author}{Chang, M.-S.}, \bibinfo{author}{Qin, Q.}, \bibinfo{author}{Zhang, W.}, \bibinfo{author}{You, L.}, \& \bibinfo{author}{Chapman, M.~S.},
			\newblock \bibinfo{title}{Coherent spinor dynamics in a spin-1 Bose condensate}.
			\newblock \emph{\bibinfo{journal}{Nature Physics}}
			\textbf{\bibinfo{volume}{1}}, \bibinfo{pages}{111--116} (\bibinfo{year}{2005}).
			
			
			
			\bibitem{Kronjager2006}%
			\bibinfo{author}{Kronj{\"a}ger, J.}, \bibinfo{author}{Becker, C.}, \bibinfo{author}{Navez, P.}, \bibinfo{author}{Bongs, K.}, \& \bibinfo{author}{Sengstock, K.},
			\newblock \bibinfo{title}{Magnetically tuned spin dynamics resonance}.
			\newblock \emph{\bibinfo{journal}{Phys. Rev. Lett.}}
			\textbf{\bibinfo{volume}{97}}, \bibinfo{pages}{110404} (\bibinfo{year}{2006}).
			
			\bibitem{Black2007}%
			\bibinfo{author}{Black, A.~T.}, \bibinfo{author}{Gomez, E.}, \bibinfo{author}{Turner, L.~D.}, \bibinfo{author}{Jung, S.}, \& \bibinfo{author}{Lett, P.~D.},
			\newblock \bibinfo{title}{Spinor dynamics in an antiferromagnetic spin-1 condensate}.
			\newblock \emph{\bibinfo{journal}{Phys. Rev. Lett.}}
			\textbf{\bibinfo{volume}{99}}, \bibinfo{pages}{070403} (\bibinfo{year}{2007}).
			
			\bibitem{Yingmei2009}%
			\bibinfo{author}{Liu, Y.}, \bibinfo{author}{Jung, S.}, \bibinfo{author}{Maxwell, S.~E.}, \bibinfo{author}{Turner, L.~D.}, \bibinfo{author}{Tiesinga, E.}, \& \bibinfo{author}{Lett, P.~D.},
			\newblock \bibinfo{title}{Quantum phase transitions and continuous observation of spinor dynamics in an antiferromagnetic condensate}.
			\newblock \emph{\bibinfo{journal}{Phys. Rev. Lett.}}
			\textbf{\bibinfo{volume}{102}}, \bibinfo{pages}{125301} (\bibinfo{year}{2009}).
			
			\bibitem{Pechkis2013}%
			\bibinfo{author}{Pechkis, H.~K.}, \bibinfo{author}{Wrubel, J.~P.}, \bibinfo{author}{Schwettmann, A.}, \bibinfo{author}{Griffin, P.~F.}, \bibinfo{author}{Barnett, R.}, \bibinfo{author}{Tiesinga, E.}, \& \bibinfo{author}{Lett, P.~D.},
			\newblock \bibinfo{title}{Spinor dynamics in an antiferromagnetic spin-1 thermal Bose gas}.
			\newblock \emph{\bibinfo{journal}{Phys. Rev. Lett.}}
			\textbf{\bibinfo{volume}{111}}, \bibinfo{pages}{025301} (\bibinfo{year}{2013}).
			
			\bibitem{Lichao2015}%
			\bibinfo{author}{Zhao, L.}, \bibinfo{author}{Jiang, J.}, \bibinfo{author}{Tang, T.}, \bibinfo{author}{Webb, M.}, \& \bibinfo{author}{Liu, Y.},
			\newblock \bibinfo{title}{Antiferromagnetic spinor condensates in a two-dimensional optical lattice}.
			\newblock \emph{\bibinfo{journal}{Phys. Rev. Lett.}}
			\textbf{\bibinfo{volume}{114}}, \bibinfo{pages}{225302}
			(\bibinfo{year}{2015}).
			
			\bibitem{He2015}%
			\bibinfo{author}{He, X.}, \bibinfo{author}{Zhu, B.}, \bibinfo{author}{Li, X.}, \bibinfo{author}{Wang, F.}, \bibinfo{author}{Xu, Z.--F.}, \& \bibinfo{author}{Wang, D.},
			\newblock \bibinfo{title}{Coherent spin-mixing dynamics in thermal Rb 87 spin-1 and spin-2 gases}.
			\newblock \emph{\bibinfo{journal}{Phys. Rev. A}}
			\textbf{\bibinfo{volume}{91}}, \bibinfo{pages}{033635} (\bibinfo{year}{2015}).
			
			
			
			\bibitem{Jiang2016}%
			\bibinfo{author}{Jiang, J.}, \bibinfo{author}{Zhao, L.}, \bibinfo{author}{Wang, S.--T.}, \bibinfo{author}{Chen, Z.}, \bibinfo{author}{Tang, T.}, \bibinfo{author}{Duan, L.--M.}, \& \bibinfo{author}{Liu, Y.},
			\newblock \bibinfo{title}{First-order superfluid-to-Mott-insulator phase transitions in spinor condensates}.
			\newblock \emph{\bibinfo{journal}{Phys. Rev. A}} \textbf{\bibinfo{volume}{93}},
			\bibinfo{pages}{063607} (\bibinfo{year}{2016}).
			
			\bibitem{Lichao2014}%
			\bibinfo{author}{Zhao, L.}, \bibinfo{author}{Jiang, J.}, \bibinfo{author}{Tang,  T.}, \bibinfo{author}{Webb, M.}, \& \bibinfo{author}{Liu, Y.},
			\newblock \bibinfo{title}{Dynamics in spinor condensates tuned by a microwave dressing field}.
			\newblock \emph{\bibinfo{journal}{Phys. Rev. A}} \textbf{\bibinfo{volume}{89}},
			\bibinfo{pages}{023608} (\bibinfo{year}{2014}).
			
			\bibitem{Lichao2018}%
			\bibinfo{author}{Zhao, L.},  \bibinfo{author}{Tang, T.},  \bibinfo{author}{Chen, Z.}, \&  \bibinfo{author} {Liu, Y.}, 
			\newblock \bibinfo{title}{Lattice-induced rapid formation of spin singlets in spin-1 spinor condensates}.
			\newblock \emph{ \bibinfo{journal}{arXiv}}:\bibinfo{pages}{1801.00773} (\bibinfo{year}{2018}).
			
			\bibitem{Jiang2014}%
			\bibinfo{author}{Jiang,  J.}, \bibinfo{author}{Zhao, L.}, \bibinfo{author}{Webb, M.}, \& \bibinfo{author}{Liu, Y.},
			\newblock \bibinfo{title}{Mapping the phase diagram of spinor condensates via adiabatic quantum phase transitions}.
			\newblock \emph{ \bibinfo{journal}{Phys. Rev. A}} \textbf{\bibinfo{volume}{90}},
			\bibinfo{pages}{023610} (\bibinfo{year}{2014}).
			
			
			\bibitem{Guan2021}%
			\bibinfo {author}{Guan, Q.} \&  \bibinfo {author}{ Lewis-Swan, R.~J.},
			\newblock \bibinfo{title}{Identifying and harnessing dynamical phase transitions for quantum-enhanced sensing}.
			\newblock \emph{\bibinfo{journal}{Phys. Rev. Research}} \textbf{\bibinfo{volume}{3}},
			\bibinfo{pages}{033199} (\bibinfo{year}{2021}).
			
			\bibitem{Zhou2023}%
			\bibinfo {author} {Zhou, L.}, \bibinfo {author} {Kong, J.}, \bibinfo {author} {Lan, Z.}, \& \bibinfo {author} {Zhang, W.}, 
			\newblock \bibinfo {title} {Dynamical quantum phase transitions in a spinor Bose-Einstein condensate and criticality enhanced quantum sensing}. 
			\newblock \emph{\bibinfo{journal}{Phys. Rev. Research}} \textbf{\bibinfo{volume}{5}},
			\bibinfo{pages}{013087} (\bibinfo{year}{2023}).
			
			\bibitem{Gerbier2021} B. Evrard, A. Qu, J. Dalibard, and F. Gerbier, Observation of fragmentation of a spinor Bose-Einstein condensate, \bibinfo{journal}{Science} \textbf{\bibinfo{volume}{373}}, \bibinfo{pages}{1340} (\bibinfo{year}{2021}).
			
			\bibitem{Chin2010}%
			\bibinfo {author} {Chin, C.}, \bibinfo {author} {Grimm, R.}, \bibinfo {author} {Julienne, P.}, \& \bibinfo {author} {Tiesinga, E.},
			\newblock \bibinfo{title} {Feshbach resonances in ultracold gases}.
			\newblock \emph{\bibinfo{journal}{Reviews of Modern Physics}} \textbf{\bibinfo{volume}{82}},
			\bibinfo{pages}{1225} (\bibinfo{year}{2010}).
			
			\bibitem{Knoop2011PRA}%
			\bibinfo{author}{Knoop, S.}, \bibinfo{author}{Schuster, T.}, \bibinfo{author}{Scelle, R.}, \bibinfo{author}{Trautmann, A.}, \bibinfo{author}{Appmeier, J.}, \bibinfo{author}{Oberthaler, M.~K.}, \bibinfo{author}{Tiesinga, E.}, \& \bibinfo{author}{Tiemann, E.},
			\newblock \bibinfo{title}{Feshbach spectroscopy and analysis of the interaction potentials of ultracold sodium}.
			\newblock \emph{\bibinfo{journal}{Phys. Rev. A}} \textbf{\bibinfo{volume}{83}},
			\bibinfo{pages}{042704} (\bibinfo{year}{2011}).
			
			
			\bibitem{Santos2007}%
			\bibinfo{author}{Santos, L.}, \bibinfo{author}{Fattori, M.}, \bibinfo{author}{Stuhler, J.}, \& \bibinfo{author}{Pfau, T.},
			\newblock \bibinfo{title}{Spinor condensates with a laser-induced quadratic Zeeman effect}.
			\newblock \emph{\bibinfo{journal}{Phys. Rev. A}} \textbf{\bibinfo{volume}{75}},
			\bibinfo{pages}{053606} (\bibinfo{year}{2007}).
			
			
			\bibitem{Leslie2009}%
			\bibinfo{author}{Leslie, S.~R.}, \bibinfo{author}{Guzman, J.}, \bibinfo{author}{Vengalattore, M.}, \bibinfo{author}{Sau, J.~D.}, \bibinfo{author}{Cohen, M.~L.}, \& \bibinfo{author}{Stamper-Kurn, D.~M.},
			\newblock \bibinfo{title}{Amplification of fluctuations in a spinor Bose-Einstein condensate}.
			\newblock \emph{\bibinfo{journal}{Phys. Rev. A}} \textbf{\bibinfo{volume}{79}},
			\bibinfo{pages}{043631} (\bibinfo{year}{2009}).
			
			\bibitem{Gerbier2006}%
			\bibinfo{author}{Gerbier, F.}, \bibinfo{author}{Widera, A.}, \bibinfo{author}{F{\"o}lling, S.}, \bibinfo{author}{Mandel, O.},  \& \bibinfo{author}{Bloch, I.},
			\newblock \bibinfo{title}{Resonant control of spin dynamics in ultracold quantum gases by microwave dressing}.
			\newblock \emph{\bibinfo{journal}{Phys. Rev. A}} \textbf{\bibinfo{volume}{73}},
			\bibinfo{pages}{041602(R)} (\bibinfo{year}{2006}).
			
			\bibitem{Bookjans2011}%
			\bibinfo{author}{Bookjans, E.~M.}, \bibinfo{author}{Vinit, A.}, \& \bibinfo{author}{Raman, C.},
			\newblock \bibinfo{title}{Quantum phase transition in an antiferromagnetic spinor Bose-Einstein condensate}.
			\newblock \emph{\bibinfo{journal}{Phys. Rev. Lett.}} \textbf{\bibinfo{volume}{107}},
			\bibinfo{pages}{195306} (\bibinfo{year}{2011}).
			
			\bibitem{Marino2022}%
			\bibinfo {author} {Marino, J.}, \bibinfo {author} {Eckstein, M.}, \bibinfo {author} {Foster, M.}, \& \bibinfo {author} {Rey, A.-M.},
			\newblock \bibinfo{title} {Dynamical phase transitions in the collisionless pre-thermal states of isolated quantum systems: theory and experiments}.
			\newblock \emph{\bibinfo{journal}{Reports on Progress in Physics}} \textbf{\bibinfo{volume}{85}}, \bibinfo{pages}{116001} (\bibinfo{year}{2022}).
			
			\bibitem{Robert2021}%
			\bibinfo{author}{ Lewis-Swan, R.~J.}, \bibinfo{author}{ Muleady, S.~R.}, \bibinfo{author}{Barberena, D.}, \bibinfo{author}{ Bollinger, J.~J.}, \& \bibinfo{author}{Rey, A.~M.},
			\newblock \bibinfo{title}{Characterizing the dynamical phase diagram of the {D}icke model via classical and quantum probes}.
			\newblock \emph{\bibinfo{journal}{Phys. Rev. Research.}} \textbf{\bibinfo{volume}{3}},
			\bibinfo{pages}{L022020} (\bibinfo{year}{2021}).
			
			\bibitem{SM} See Supplemental Materials for more details of experimental procedures, data analysis, and theoretical calculations.	
			
			\bibitem{Fujimoto2019}
			\bibinfo{author}{Fujimoto, K.} \& \bibinfo{author}{ Uchino, S.}, \bibinfo{title}{ Floquet spinor Bose gases}. \emph{\bibinfo{journal}{Phys. Rev. Research.}} \textbf{\bibinfo{volume}{1}}, \bibinfo{pages}{033132} (\bibinfo{year}{2019}).
			
			\bibitem{Zhang2023}
			\bibinfo{author}{Zhang, Y.}, \bibinfo{author}{ Chen, Y.}, \bibinfo{author}{ Lyu, H.},\& \bibinfo{author}{Zhang, Y.},  \bibinfo{title}{Quantum phases in spin-orbit-coupled Floquet spinor Bose gases}. \emph{\bibinfo{journal}{Phys. Rev. Research.}} \textbf{\bibinfo{volume}{5}}, \bibinfo{pages}{023160} (\bibinfo{year}{2023}).
			
			\bibitem{Evrard2019}
			\bibinfo{author}{Evrard, B.}, \bibinfo{author}{ Qu, A.}, \bibinfo{author}{Jim\'{e}nez-Garc\'{i}a, K.}, \bibinfo{author}{Dalibard, J.}, \& \bibinfo{author}{Gerbier, F.}, \bibinfo{title}{ Relaxation and hysteresis near Shapiro resonances in a driven spinor condensate}. \emph{\bibinfo{journal}{Phys. Rev. A}} \textbf{\bibinfo{volume}{100}}, \bibinfo{pages}{023604} (\bibinfo{year}{2019}).
			
			\bibitem{Li2019}
			\bibinfo{author}{Li, Z.-C.}, \bibinfo{author}{Jiang, Q.-H.}, \bibinfo{author}{Lan, Z.}, \bibinfo{author}{Zhang, W.},\& \bibinfo{author}{Zhou, L.}, \bibinfo{title}{ Nonlinear Floquet dynamics of spinor condensates in an optical cavity: Cavity-amplified parametric resonance}. \emph{\bibinfo{journal}{Phys. Rev. A}} \textbf{\bibinfo{volume}{100}}, \bibinfo{pages}{033617} (\bibinfo{year}{2019}).
			
			\bibitem{Jose2023}%
			\bibinfo{author}{Jose, S.~M.},  \bibinfo{author}{Sah, K.}, \&  \bibinfo{author} {Nath, R.}, \bibinfo{title}{Patterns, spin-spin correlations and competing instabilities in driven quasi-two-dimensional spin-1 Bose-Einstein condensates}. \emph{ \bibinfo{journal}{arXiv}}:\bibinfo{pages}{2305.12385} (\bibinfo{year}{2023}).
			
		}
	\end{thebibliography}
\end{document}